# Origin of the 60º and 90º dislocations and their role in strain relief in lattice-mismatched heteroepitaxy of fcc materials

Atul Konkar
Nanostructure Materials & Devices Laboratory
Department of Materials Science
University of Southern California
Los Angeles, CA 90089

**Abstract:**
Strain relief in lattice mismatched heteroepitaxy is mediated by formation and/or propagation of dislocations. Due to their technological significance, the process of strain relief in materials with face-centred cubic (fcc) lattices has been analyzed by several researchers[1,2] following the work by Matthews and co-workers in the late 1960s to early 1970s[3-6]. In the Matthews model, it is assumed that the strain relieved by *any* misfit dislocation is equal to the edge component of the dislocation burgers vector in the interface plane. This assumption has been used in all subsequent analyses of strain relief in lattice mismatched heteroepitaxy[1,2]. Based upon the known three-dimensional atomic structure of the dislocations in fcc lattices, we show that the assumption is not valid for the 60º dislocations that form/expand via the conservative glide process. For compressively (tensilely) strained films the assumption is valid only for the 90º dislocations that form/expand via aggregation of vacancies (interstitials).

Lattice-mismatched heteroepitaxy or strain-layer epitaxy, in which a material of a different bulk lattice constant than that of the substrate is deposited in the form of a film, plays a central role in many of the InGaAs/GaAs, InGaAsP/InP, and the emerging SiGe/Si based electronic and optoelectronic devices. The strain is used as an additional degree of freedom to engineer the electronic structure of the active layers. The lattice mismatch strain leads to formation/expansion/motion of dislocations in such structures. The most commonly observed dislocations in fcc strained systems, such as the ones noted above, have their misfit dislocations (i.e. part of the dislocation line that lies in the film/substrate interface plane) along the orthogonal [110] and [$\bar{1}$10] directions. These dislocations are conventionally categorized into two groups: 60º and 90º (also known as edge-type) dislocations, depending upon the angle between the misfit dislocation line and its burgers vector ($\bar{b}$). The dislocations impact the properties of the film, in most cases adversely, and therefore understanding and control of dislocations is an important objective in the synthesis of such films. The first ground state thermodynamic analysis at 0K of lattice-mismatched heteroepitaxy was done for one monolayer (ML) thick film[7], subsequently such analysis was extended to films with finite (>> 1 ML) thick films[8]. The analysis in references 7 and 8 was limited to the case of infinitely long edge-type misfit dislocations. Subsequently, such analysis was extended by Matthews and co-workers[3-6] to dislocations that have a finite extent in the interface plane and are of mixed character. In their analysis the strain relieved by the misfit dislocation segment was assumed to be equal to the projection of the edge component of the burgers vector on to the interface plane[9]. We have examined the validity of this assumption (which we refer to as Matthews assumption) based on the known three-dimensional structure of dislocations in fcc lattices. We show that the Matthews assumption is valid only for 90º dislocations that form/expand via vacancy/interstitial aggregation and we show that it is not applicable to the 60º dislocation that form or expand via the conservative glide process.

Following the typical experimental situation, we consider a film of material F with a bulk lattice parameter, $a_f^0$, on the (001) surface of a semi-infinite perfect substrate of material S with a bulk lattice parameter, $a_s^0$. Both materials are assumed to have fcc lattice structure in their bulk form. A schematic of the system under consideration is shown in fig. 1(a). We assume $a_f^0 > a_s^0$, so that the film is compressively strained when it is coherent with the substrate. An equivalent description of coherency is that the spacing of the (110) and ($\bar{1}$10) film planes is same as the corresponding planes of the substrate. Figure 1(b) shows a schematic of (110) cross-section of the lattice of the film/substrate system. In this analysis we do not consider infinitely long dislocations that span the entire film/substrate interface. This condition is consistent with the typical experimental situation where the dislocation velocities (glide or climb) and the time scales involved are such that the dislocation can expand at most ~ 1 mm, whereas the typical substrate size is ~ 10 - 1000 mm. The three dimensional structure of the dislocations considered here is that of dislocation half-loops, i.e. dislocations that have three-segments, one segment lying at the interface (misfit dislocation), and the other two segment that span the film thickness (threading dislocations) and terminating at the film surface.

Strain relief can be considered to be increase in the spacing of the [110] and [$\bar{1}$10] planes of the film from the initial value that is equal to the spacing of the corresponding planes of the substrate (coherently strained film) to their value in the bulk material F (fully relaxed film). To illustrate the process of strain relief via dislocations we first consider the case of 90º dislocations. Shown in figure 2(a) and (b) are the three-dimensional schematic of such a dislocation half-loop and (110) cross-section of the lattice across the misfit segment, respectively. The cross-section shows the familiar missing half-plane picture of edge-type dislocations. The dislocation burgers vector is $1/2a_s^0[\bar{1}10]$ and its magnitude corresponds to the width of the missing half-plane. If the average spacing of the 90º dislocations in the [110] and [$\bar{1}$10] directions is the same and is equal to d, then from geometry it follows that the relieved strain [2], (f−ε), is

$$f-\varepsilon = |1/2a_s^0[110]|/d = |\bar{b}|/d \quad \text{-----------------------------} \quad (1)$$

where $|\bar{b}|$ is the magnitude of the burgers vector for the 90º dislocations.

A comparison of fig. 2(a) and 2(b) implies that the 90º dislocation half-loop may be considered as a disc of vacancies with thickness $|1/2a_s^0[110]|$ and is analogous to Frank partial bounding an intrinsic stacking fault in fcc materials[10] that is familiar to most materials scientists. The difference is that the Frank partials have burgers vector $1/3a[111]$, which is not a perfect lattice vector, whereas the 90º dislocations have a burgers vector that is a perfect lattice vector. Similar to the Frank partials the 90º dislocations are edge-type along the entire dislocation line, a geometrically necessary requirement. Indeed, any dislocation loop (or half-loop) that can be constructed as a vacancy disc will have an edge character along its periphery and will relieve strain. We refer to all such dislocations as dilational dislocations. We note that dilational dislocations different from the typical 90º dislocations have been experimentally observed[11]. These dilational dislocation loops are nucleated on a pre-existing 60º dislocations, and have a burgers vector of type $1/2a_f[101]$[12] and the misfit segment is along the [100] direction. The magnitude of the strain relieved by any dilational dislocations can be shown to be:

$$f-\varepsilon = |\bar{b}_{eff}|/d \quad \text{--------------------------------------------------} \quad (2)$$

where $|\bar{b}_{eff}|$ is the projection of the burgers vector in the interface plane and d is the dislocation spacing. The amount of strain relief by dilational dislocations is consistent with the Matthews assumption. We remark here that unambiguous determination of a dislocation loop (or half-loop) to be dilational type can be made only by analyzing its three-dimensional structure. This determination can not be made by analyzing only the misfit segment. The above analysis also applies for $a_f^0 < a_s^0$ (tensile film stress) and in this case the strain will be relieved by dilational dislocations that form/expand via aggregation of interstitials.

A schematic three-dimensional structure of 60º dislocation half-loop and (110) cross-section of the lattice across the misfit segment are shown in figure 3 (a) and (b), respectively. The entire half-loop lies in the (1$\bar{1}$1) plane which is the natural glide plane for fcc lattices. The dislocation half-loop can be considered to be shear displacement of atoms enclosed by the dislocation line with a displacement vector of $1/2a_f[101]$. Note that a geometrically necessary surface step is concomitantly created. The cross-section (see fig. 3(b)) shows that the spacing of the ($\bar{1}$10) film planes is not affected by such a dislocation. The only change is a line of vacancies along the misfit segment. Therefore it follows that the 60º dislocation does not relieve strain through the thickness of the film and the Matthews assumption is not satisfied by such dislocations. To the first order, for thick films (thickness >> burgers vector), we can assume that the 60º dislocations do not relieve any strain.

We have shown that the origin of the 60º and 90º dislocations and their role in strain relief is entirely different. We have shown that only those dislocations that form/expand via vacancy/interstitial aggregation result in strain relief. In contrast, dislocations that form/expand via glide do not relieve strain. The strain relief will be controlled by the processes of vacancy/interstitial formation/capture and their subsequent diffusion. These aspects have received minimal consideration in the literature so far and should be examined in detail to provide better understanding of the strain relief process. Though the results have been illustrated specifically for fcc lattices they are general and applicable to other lattice systems.

**Acknowledgements**
This work was supported by AFOSR/DARPA under the DoD Defense University Research Initiative on NanoTechnology program Grant No. AFOSR F49620-01-1-0474. I thank Prof. Anupam Madhukar for the thought provoking discussions and providing inspiration for this work.

**References:**

[1] For recent review of strained epitaxy in III-V materials see: Dunstan, D. J., Strain and strain relaxation in semiconductors. *J. Mater. Sci: Mater. in Electron.* **8,** 337 (1997). In the context of this paper, section 4 is relevant.

[2] For recent review of SiGe-Si system see: Bolkhovityanov, Yu. B., Pchelyakov, O. P., and Chikichev, S. I., Silicon-germanium epilayers: physical fundamentals of growing strained and fully relaxed heterostructures. *Phys.-Uspekhi* **44,** 655, (2001). Section 2 is the most relevant. See sub-section 2.2.1 for definition of relieved strain used in the text. A simple derivation of the energy-balance model for analyzing strain-relief process is also given in this sub-section. The Matthews assumption is incorporated in equation (4).

[3] Matthews, J.W., Accommodation of misfit across the interface between single-crystal films of various face-centred cubic metals. *Phil. Mag.* **13**, 1207 (1966);

[4] Matthews, J. W. and Crawford, J. L., Accommodation of misfit between single-crystal films of nickel and copper. *Thin Solid Films* **5**, 187 (1970);

[5] Matthews, J. W. and Blakeslee, A. E., Defects in epitaxial multilayers I. Misfit dislocations. *J. Cryst. Growth* **27**, 118 (1974);

[6] Matthews, J. W., Defects associated with the accommodation of misfit between crystals. *J. Vac. Sci. Technol.* **12**, 126 (1975).

[7] Frank, F. C. and Van der Merwe, J. H., One-dimensional dislocations. II. Misfitting monolayers and oriented overgrowth. Proc. *Roy. Soc. London Ser A* **198**, 216 (1950).

[8] Van Der Merwe, J. H., Crystal Interfaces. Part II. Finite Overgrowths. *J. Appl. Phys.* **34**, 123 (1963).

[9] We have researched the origin of this assumption and to the best of our knowledge the first instance of this assumption in text form occurs in reference 2, p. 1214, last paragraph: "The dislocation labeled A and B are therefore mixed dislocations. The misfit that they can accommodate is equal to their edge component projected into the film plane." The assumption was incorporated in mathematical form in their subsequent papers, as an example see equation (3) in reference 4. In this equation the factor cosλ captures the Matthews assumption.

[10] Hirth, J. P. and Lothe, J., *Theory of Dislocations.*, Keieger Publishing Company, Malabar, Florida, USA, (1992).

[11] Liu, X. W., Hopgood, A. A., Usher, B. F., Wang, H., and Braithwaite, N. St. J., Formation of misfit dislocations in strained-layer GaAs/InxGa1-xAs/GaAs heterostructures during postfabrication thermal processing. *J. Appl. Phys.* **94**, 7496 (2003).

[12] The burgers vector is given in units of $a_f$ which refers to appropriate average of the instantaneous lattice parameters of the film along the three orthogonal <100> directions. Such averaging is required since the film may have orthorhombic symmetry in the case of non-uniform strain relaxation in the orthogonal in-plane directions.

**Figure captions:**

Figure 1 (a) Schematic of the system under consideration. (b) Schematic (110) cross-section of the lattice showing coherently strained film. Two adjacent (110) planes separated by ¼ [110] are projected, the circles filled with dark colors are lattice points lie in one (110) plane and the light circles are lattice points from the adjacent (110) plane.

Figure 2 (a) Schematic of the three dimensional structure of 90º dislocation half-loop. The entire half-loop lies in the ($\bar{1}$10) plane. (b) Schematic (110) cross-section of the lattice across 90º misfit dislocations. The boxes with dotted lines represent the 'missing half-plane' of dislocations.

Figure 3 (a) Schematic of the three dimensional structure of 60º dislocation half-loop. The entire half-loop lies in the (1$\bar{1}$1) glide plane. (b) Schematic (110) cross-section of the lattice across 60º misfit dislocations. The unfilled circles represent locations from which atoms have moved due to dislocation glide on the (1$\bar{1}$1) planes. The burgers vector, ½ $a_f$[$\bar{1}$01], connects the dark green circle to the adjacent light green circle on the (1$\bar{1}$1) plane and lies out of the plane of the paper.

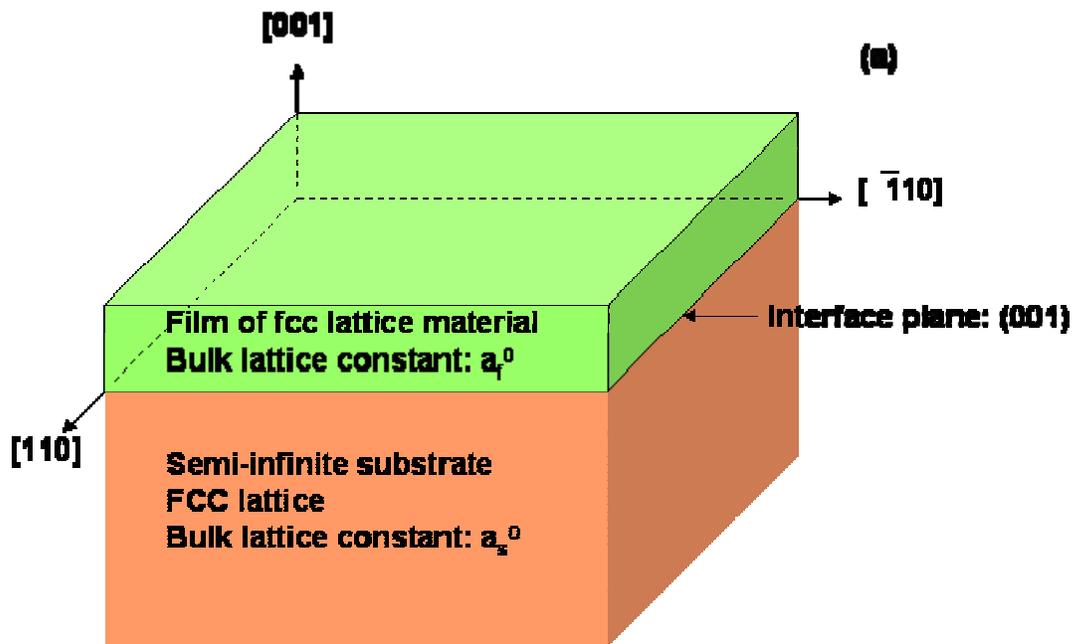

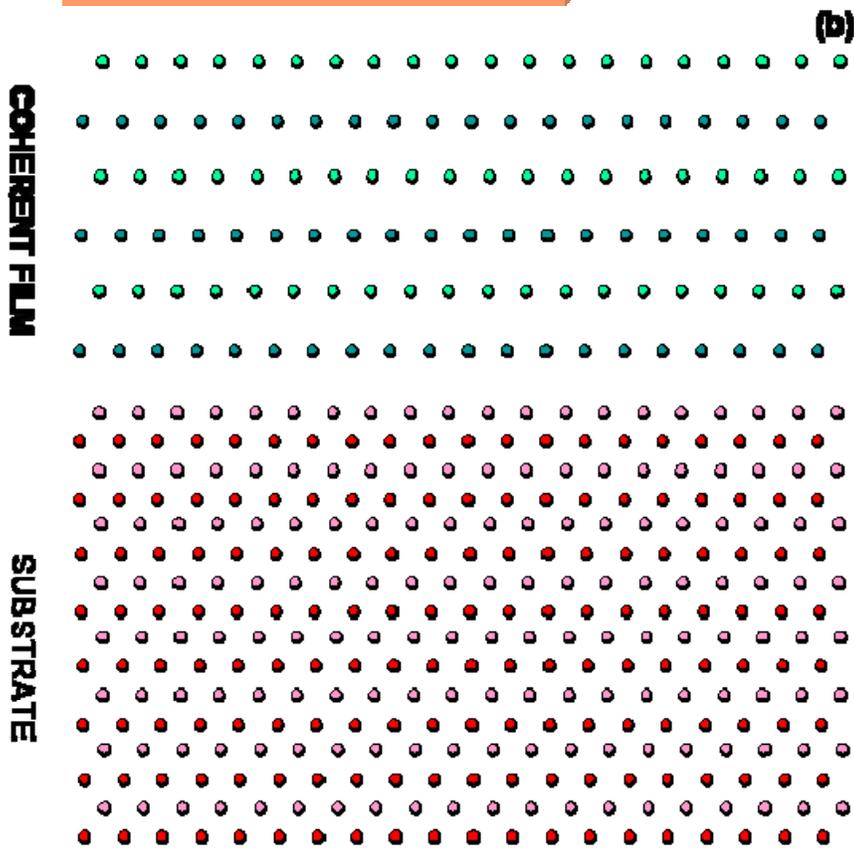

Konkar_Fig1

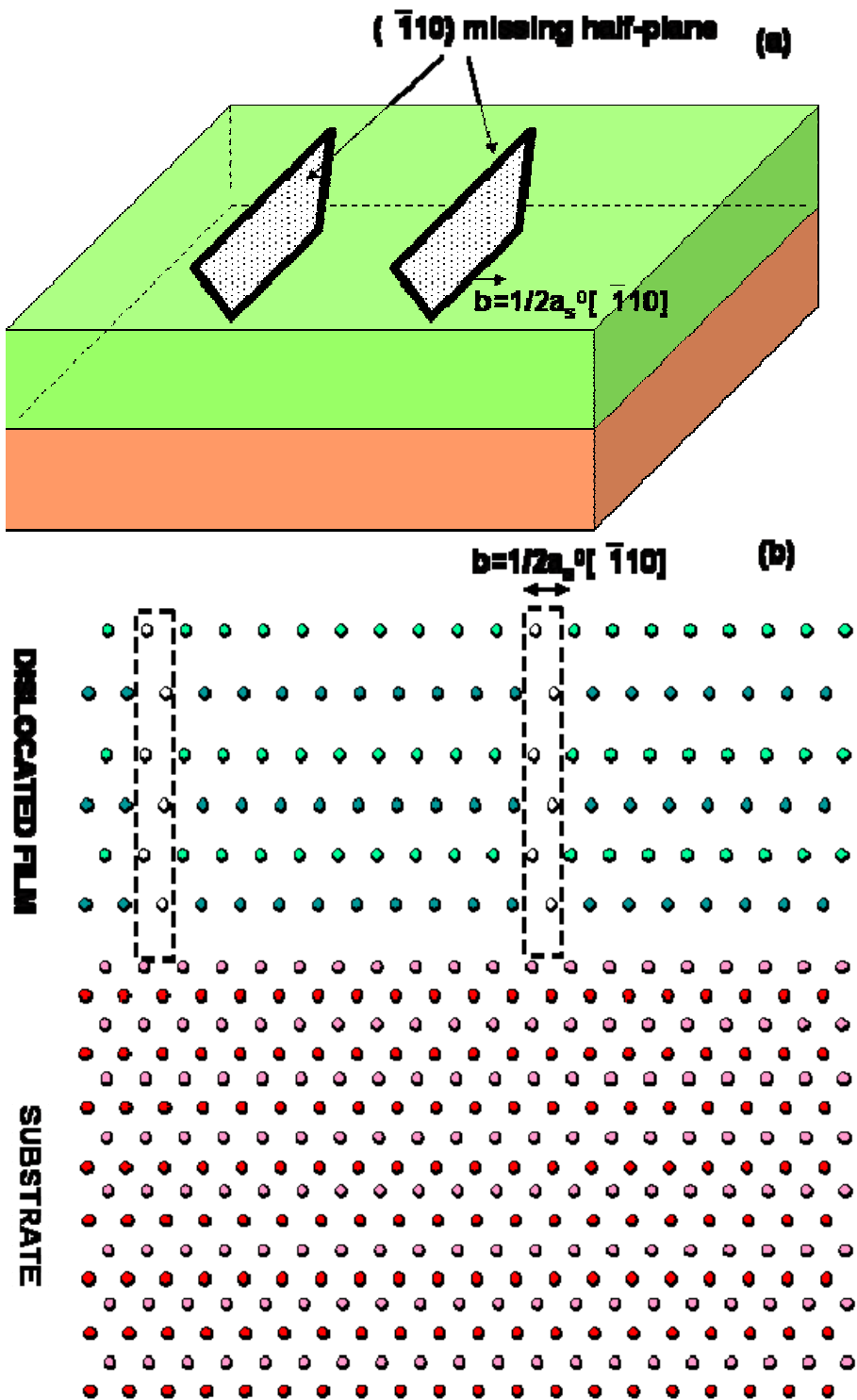

Konkar_Fig2

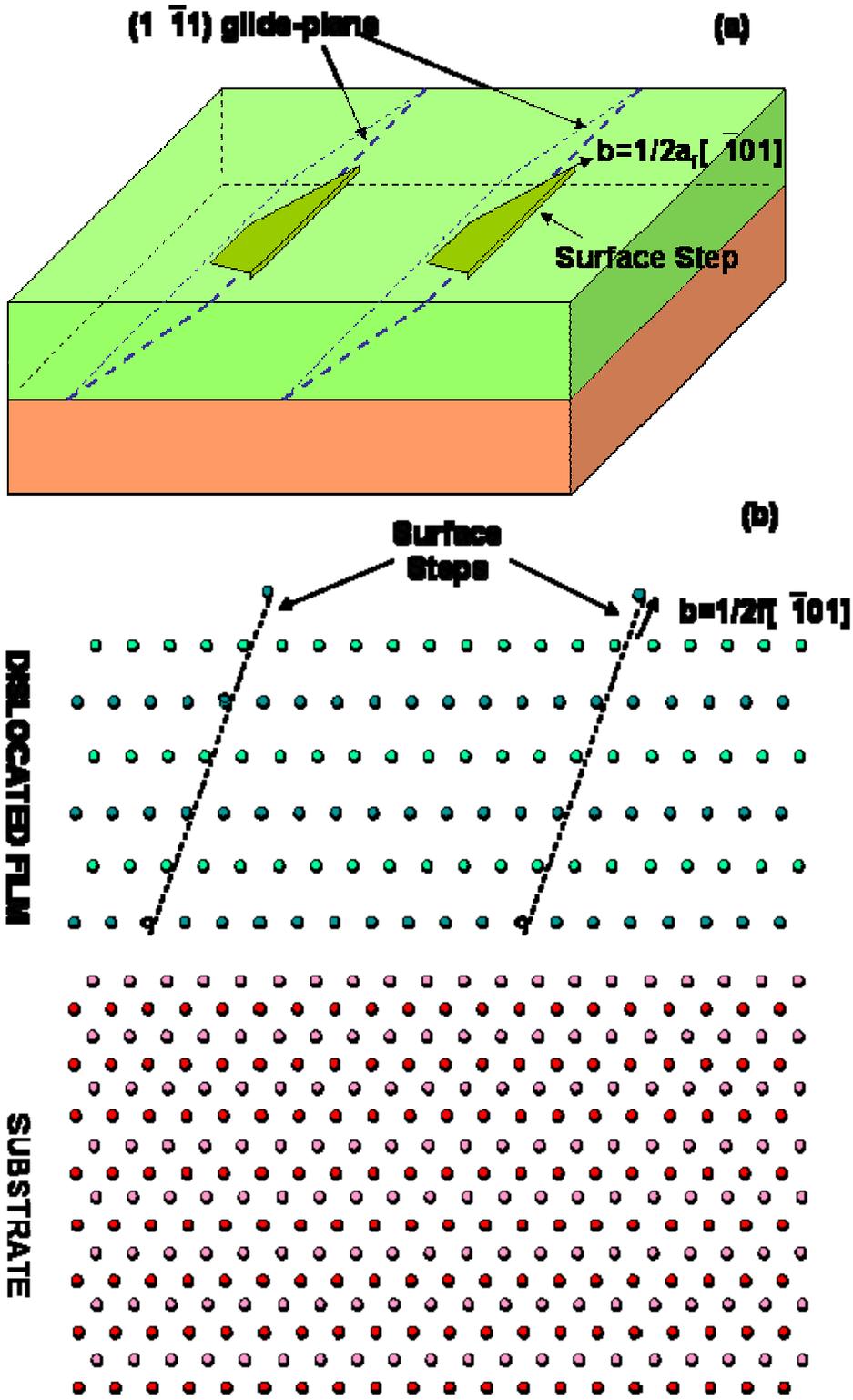

Konkar_Fig3